\documentclass[dvips,preprint]{imsart}
\RequirePackage[OT1]{fontenc}
\RequirePackage{amsthm,amsmath,natbib}
\RequirePackage[colorlinks]{hyperref}
\RequirePackage{hypernat}
\usepackage{lscape}
\RequirePackage{epsf}
\usepackage{epsfig}
\usepackage{rotating}

\startlocaldefs
\numberwithin{equation}{section}
\theoremstyle{plain}

\endlocaldefs

\begin{document}

\begin{frontmatter}
\title{A Family of Generalized Beta Distributions for Income}

\begin{aug}
\author{\fnms{J. H.} \snm{Sepanski}\corref{}}
\address{Department of Mathematics\\
Central Michigan University\\
Mount Pleasant, MI 48859\\
email: sepan1jh@cmich.edu
}
\end{aug}

\begin{aug}
\author{\fnms{Lingji} \snm{Kong}}

\address{Liaoning Normal University\\
LNU-MSU College of International Business\\
P.O.Box 13001\\
850 Huanghe Road\\
Dalian, Liaoning 116029\\
China\\}

\end{aug}

\begin{abstract}
The mathematical properties of a family of generalized beta distribution, including beta-normal, skewed-t, log-F, beta-exponential, beta-Weibull distributions have recently been studied in several publications. This paper applies these distributions to the modeling of the size distribution of income and computes the maximum likelihood estimation estimates of parameters.  Their performances are compared to the widely used generalized beta distributions of the first and second types in terms of measures of goodness of fit.
\end{abstract}

\begin{keyword}[class=AMS]
\kwd[Primary ]{62P20} \kwd{62F07}
\end{keyword}

\begin{keyword}
\kwd{generalized beta distribution, skewed-t distribution, log-F distribution, Weibull distribution}
\end{keyword}
\tableofcontents
\end{frontmatter}

\section{Introduction}

Consider the distribution function of a beta random variable given by
\begin{equation}
G(y)=[B(\alpha ,\beta )]^{-1} \int _{0}^{y}t^{\alpha -1} (1-t)^{\beta -1} dt, \label{cs}  \end{equation}
for $0<y<1,$ where $\alpha >0, \beta >0$ and the beta function $B(\alpha ,\beta )=\Gamma (\alpha +\beta )/\left[\Gamma (\alpha )\Gamma (\beta )\right]$.  Note the domain of $G(\cdot )$ is (0, 1). Use the fact that the range of a cumulative distribution function (cdf) is (0, 1), replacing the upper limit \textit{y} of the integration in (\ref{cs}) with a cdf $F(\cdot )$ has been studied in several papers reviewed below. The resulting probability density function is
\begin{equation}
g_{F} (x)=\left[B(\alpha ,\beta )\right]^{-1} f(x)\left[F(x)\right]^{\alpha -1} \left[1-F(x)\right]^{\beta -1}, \label{cs1}
\end{equation}
where $f(\cdot )$ is the derivative of $F(\cdot )$and therefore is the corresponding probability density function if $F(\cdot )$ is a distribution function. For simplicity, this distribution will be called the generalized beta-\textit{F }distribution hereafter.

Jones (2004) introduced this as the probability density function of the transformed random variable $X=F^{-1} (Y)$ where $Y$is a Beta random variable with parameters of $\alpha $ and $\beta .$ The density function form in (\ref{cs1}) was also alternatively described as a simple generalization of the use of the collection of order statistics distributions associated with \textit{F}.  Jones (2004) and Ferreira, etc. (2004) explored general properties of this family of distributions and examined the special cases of skewed-t and log-F distributions. Since $F(\cdot )$ can be any distribution function, the family of this generalized beta-\textit{F} distribution is a very rich one and can be further explored. This family of distribution was first introduced by Singh et al. (1988) and has since been studied for several distribution functions. In this paper, it is applied to the analyses of placecountry-regionU.S. family income data.

Numerous distributions (see McDonald, 1984, and references therein), including gamma, beta, Singh-Maddala (or Burr), Pareto, Weibull, and generalized beta of first and second kinds, have been used to model the size distribution of income.  McDonald (1984) fit the above models to the income data of 1970, 1975, and 1980 and concluded that the generalized beta of the second type provided the best relative fit and that the Singh-Maddala (SM) distribution provided a better fit than the generalized beta of the first kind. McDonald also discussed the relationships between several widely used models for the income distribution, those relationships can be expanded to the family of the generalized beta-\textit{F} distribution in (\ref{cs1}) that includes some of the distributions as its special cases.

In this paper, examples of the family of the generalized beta-\textit{F} distribution described in (\ref{cs1}) in existing literature are summarized in section 2. The distributions tabulated in Table 1 are fit to the U.S. family income data presented in a grouped format on the website of the Census Bureau. Outlines of the maximum likelihood estimation of unknown population parameters involved in the generalized beta-\textit{F} distribution function and in the $F(\cdot )$function are derived for the grouped income data in section 3.  The equations to be maximized and the gradients do not have closed forms and depend on the function $F(\cdot )$ of interest. Besides the parameter estimates and associated estimated value for the mean, goodness-of-fit values including sum of the squared errors, sum of the absolute deviations and chi-squares are reported for comparisons in section 4.  The performance comparisons of the distributions are also presented.

\section{The Models}

The probability functions of interest and their means and moments are summarized in Table 1 in this section. Technical details on the characteristics such as the shapes, moments, skewness, and limiting distribution as some parameters tend to extreme values of each distribution can also be found in the provided reference.

The generalized beta of the first (GB1) and second (GB2) kind (McDonald, 1984) are respectively defined by
\begin{equation}
g(y)=\frac{ay^{a\alpha -1} [1-(y/b)^{a} ]^{\beta -1} }{b^{a\alpha } B(\alpha ,\beta )} ,{\rm \; }0\le y\le b, \label{tp1}
\end{equation}
\begin{equation}
g(y)=\frac{ay^{a\alpha -1} }{b^{a\alpha } B(\alpha ,\beta )[1+(y/b)^{a} ]^{\alpha +\beta } } ,{\rm \; }0\le y.\label{tp2}
\end{equation}
They are special cases of the generalized beta-\textit{F} distribution with $F(x)=(x/b)^{a} $ and $F(x)=1-[1+(x/b)^{a} ]^{-1} $ for $x>0$ in (\ref{cs1}), respectively.  The underlying distribution of income in Thurow (1970) with \textit{a=1} in (\ref{tp1}) is therefore also a special case with a distribution function \textit{F} of a uniform distribution over the interval $(0,b)$. The Singh-Maddala distribution with a density function of $a\beta y^(a-1)/[1+(y/b)^{a} ]^{\beta+1} $ is a special case of generalized beta of the second kind with the beta parameter $\alpha =1.$ Note that these distributions are unimodal.

Eugene (2002)studies the properties of a beta-normal (BN) distribution, i.e., \textit{F} is a normal distribution function. Gupta and Nadarajah (2004) further derived a different form of the moments of the beta-normal distribution. The beta-normal can be both bimodal and unimodal. Eugene (2002) showed that it is skewed to the right when $\alpha >\beta $ and skewed-to the left when $\alpha <\beta $. When $\alpha =\beta $, it is symmetric about$\mu $. It has heavy symmetric tails when $\alpha <1$and $\beta <1$ in which bimodality eventually occurs as $\alpha $(=$\beta $) decreases. Also when $\alpha >1$ and$\beta >1$, it has long symmetric tails with a higher peak associated with a larger value of $\alpha $(=$\beta $).

The two particular distributions that Jones (2004) believed to provide the most tractable instances of families with power and exponential tails are the skew-t distribution (Beta-\textit{T}) and the log-F (Beta-Logistic) distribution. The skew-t distribution (Jones, 2001, and Jones and Faddy, 2003) can be derived with $F(t)=[1+t(\sqrt{a+b+t^{2} } )^{-1} ]/2,$ which is the distribution function of the scaled student t distribution on 2 degrees of freedom, with scaling factor$\sqrt{(a+b)/2} $.  The skewed-t reduces to the symmetric Student's t distribution when \textit{a=b} and becomes skewed when \textit{a$\ne $b}.  It is unimodal and heavy tailed, and the skewness measured based on the third moment is a monotone increasing function of \textit{a }for fixed \textit{b} and a monotone decreasing function of \textit{b} for fixed \textit{a}.

The log-F distribution is a special case of family (\ref{cs1}) with the standard logistic distribution \textit{F(x)=$e^{x} /(1+e^{x} )$ } which can also be other types of generalized logistic distributions. Brown \textit{et al}. (2002) presented examples of application areas including survival analyses in which log-normal, Weibull, log-logistic, and generalized gamma was shown to be special cases of the log-F model; see Kalbfleish and Prentice (1980).    The log-F is unimodal and can be symmetric, or skewed-to the left or to the right. A generalized four-parameter version of log-F with location parameter \textit{a}, scale parameter \textit{b}, shape parameters $\alpha $and replacing \textit{x} by $(x-a)/b$ in the distribution function \textit{F(x)} will be fit to the income data in this paper.

   Nadarajan (2004)derived the moment generating function, skewness, kurtosis
and other properties for the beta-exponential (BE) distribution with an exponential \textit{F}. Both measures of skewness and kurtosis are shown to decrease monotonically with the parameters $\alpha $ and $\beta .$ Famoye, etc. (2005) studied the beta-Weibull distribution with \textit{F(x)=}1-exp(-$ax^{b} $).  The beta-Weibull is unimodal, and the mode is at the point of 0 when \textit{b $<$ 1}. That is, beta-Weibull distribution (BW) has a inversed-J shape when \textit{b $<$ 1}.  Note that the exponential distribution is a special case of Weibull distribution. Nadarajan and Ktoz (2004) investigated the unimodal beta-Gumbel distribution in the hope of attracting wider applicability in engineering due the wide applications of the Gumbel distribution in the field and showed that it has a single mode and an increasing hazard function. \textbf{}

The following Table 1 lists moments and means for various beta-\textit{F} distributions to be fit to the size distribution of income. The means in the table will be calculated as a check of the validity of the parameters produced from computer algorithms in the next section. Let $\Phi (x;\mu ,\sigma )$ be the distribution function of a normal random variable with mean $\mu $ and standard deviation $\sigma $ and digamma $\Psi (x)=d\log \Gamma (x)/dx$be the Euler's psi function; see Gradshteyn and Ryzhik (2000)  Define $I_{n,k} $ $=\int _{-\infty }^{\infty }x^{n} f(x)(1-F)^{k}  dx$


{
\begin{sidewaystable}
\hspace{-1in}
\caption{Distributions and Their Moments}
\begin{tabular} {|p{0.7in}|l|l|p{2.5in}|}\hline
Model & \textit{F(x)} in (\ref{cs1}) & Moments $E(X^{n} )$ & Mean \\ \hline
GB1 & $(x/b)^{a} $ & $\frac{b^{n} B(\alpha +\beta ,n/a)}{B(\alpha ,n/a)} $ & $\frac{b^{} B(\alpha +\beta ,1/a)}{B(\alpha ,1/a)} $ \\ \hline
GB2 & $1-\frac{1}{1+(x/b)^{a} } $ & $\frac{b^{n} B(\alpha +n/a,\beta -n/a)}{B(\alpha ,\beta )} $ & $\frac{bB(\alpha +1/a,\beta -1/a)}{B(\alpha ,\beta )} $ \\ \hline
{Beta-Normal}\newline (BN)  & $\Phi (x;\mu ,\sigma )$ & $\begin{array}{l} {\mu ^{n} +\frac{\mu ^{n} }{B(\alpha ,\beta )} \sum _{j=0}^{\beta -1}(-1)^{j} \big\{ \left(_{j}^{\beta -1} \right)}\\{ \; \times {\displaystyle\sum _{i=1}^{n}(-1)^{j} } \left(_{i}^{n} \right)\left(\frac{\sigma }{\mu } \right)^{i} } \\ {\;  \times {\displaystyle\sum _{k=0}^{\alpha +j-1}}(-1)^{k}  \left(_{k}^{\alpha +j-1} \right)I_{i,k} +(-1)^{i} I_{i,\alpha +j-1}} \end{array}$ & $\begin{array}{l} {\frac{\sigma }{B(\alpha ,\beta )} {\displaystyle\sum _{j=0}^{\beta -1}}\{{\displaystyle \sum _{k=0}^{\alpha +j-2}} \frac{(-1)^{j+k} }{k+1}  \left(_{j}^{\beta -1} \right)\left(_{k}^{\alpha +j-1} \right)\delta _{k+1} } \\ {{\rm \; \; \; }+\frac{(-1)^{\alpha } -(-1)^{j} }{\alpha +j} \left(_{j}^{\beta -1} \right)\left(_{k}^{\alpha +j-1} \right)\delta _{k+1} \big\}+\mu  } \end{array}$ \\ \hline
Skew-t & $\begin{array}{l} {\frac{1}{2} \left(1+\frac{x}{\sqrt{a+b+x^{2} } } \right)}, \\ a=\alpha,  b=\beta  \end{array}$ & $\begin{array}{l} {\frac{(a+b)^{n/2} }{2^{n} B(a,b)}
{\displaystyle\sum _{i=0}^{n}} \left(_{i}^{n} \right)(-1)^{i} B\left(a+\frac{n}{2} -i,b-\frac{n}{2} +i\right)}\\{{\rm for \; \;} a>n/2,b>n/2.}\end{array}$ & $\frac{(a-b)\sqrt{a+b)} }{2} \frac{\Gamma (a-1/2)\Gamma (b-1/2)}{\Gamma (a)\Gamma (b)} $ \\ \hline
Log-F & $\displaystyle{\frac{e^{x} }{1+e^{x} }} $ & $\begin{array}{l} {\frac{1}{B(\alpha ,\beta )}{ \displaystyle\sum _{j=0}^{\beta -1}(-1)^{j}} \left(_{j}^{\beta -1} \right)}\\ {\; \; \times\left\{ \displaystyle{\sum _{k=0}^{\alpha +j-1}(-1)^{k} }\left(_{k}^{\alpha +j-1} \right) I_{n,k}\right\}}\end{array}$  & $\begin{array}{l}{\frac{1}{B(\alpha ,\beta )}  {\displaystyle \sum _{j=0}^{\beta -1}(-1)^{j}} \left(_j^{\beta -1}\right)} \\ {\;\; \times {\displaystyle\sum _{k=0}^{\alpha +j-1}(-1)^{k}} \left(_k^{\alpha +j-1} \right) I_{1,k}}\end{array} $ \\ \hline
Beta-exponential\newline (BE) & 1-exp(-$ax$) & $\frac{(-1)^n}{\lambda^n B(\alpha, \beta)} \frac{\partial^n}{\partial p^n}B\left(\alpha,\alpha+p+1\right)\vert_{p=\alpha+\beta-1} $ & $\frac{\Psi (\alpha +\beta )-\Psi (\beta )}{a} $ \\ \hline
Beta-Weibull\newline (BW) & 1-exp(-$ax^{b} $) & $\frac{\Gamma (\alpha +\beta )\Gamma (n/b+1)}{\Gamma (\beta )a^{n/b} }{\displaystyle \sum _{k=0}^{\infty }}\frac{(-1)^{k} (\beta +k)^{-(n+b)/b} }{k!\Gamma (\alpha -k)}  $ & $\frac{\Gamma (\alpha +\beta )\Gamma (1/b+1)}{\Gamma (\beta )a^{1/b} } {\displaystyle \sum _{k=0}^{\infty }}\frac{(-1)^{k} (\beta +k)^{-(1+b)/b} }{k!\Gamma (\alpha -k)}  $ \\ \hline
\end{tabular}
\end{sidewaystable}

}

\newpage

\section{Maximum Likelihood Estimation}

The income data were in a grouped format with only the frequency and mean income of each group given.  Let \textit{G} and \textit{g }be the respective cumulative distribution and probability density function of a beta random variable $Y$as in (\ref{cs}). Let $\theta _{G} $=$(\alpha ,\beta )^{T} $and $\theta _{F} $=$(a,b)^{T} $be the column vectors of parameters associated with the beta distribution \textit{G} and the distribution function \textit{F} in (\ref{cs}) and (\ref{cs1}), respectively.  Define the probability
\begin{equation}
  P_{i} (\theta _{G} ,\theta _{F} )=\int _{I_{i} }g_{F}  (x;\theta _{G} ,\theta _{F} )dx =[B(\alpha ,\beta )]^{-1} \int _{F(x_{i-1} )}^{F(x_{i} )}t^{\alpha -1} (1-t)^{\beta -1} dt . \label{hp1}
\end{equation}
 It is the proportion of the population in the \textit{i}th of the \textit{r} income groups defined by the interval$I_{i} $$=[x_{i-1} ,x_{i} )$.  The likelihood function for the data is therefore given by

\[N!\mathop{\Pi }\limits_{i=1}^{r} \frac{[P_{i} (\theta _{G} ,\theta _{F} )]^{n_{i} } }{n_{i} !} \]

where $n_{i} $, $i=1,\cdots ,r,$ is the frequency of the \textit{i}th group and $N=\sum _{i=1}^{r}n_{i}  $. The maximum log-likelihood estimators are obtained by maximizing
\begin{equation}
L(\theta _{G} ,\theta _{F} )=\sum _{i-1}^{r}n_{i} \ln  P_{i} (\theta _{G} ,\theta _{F} ). \label{hp2}
\end{equation}

It is well known that the resulting estimators by maximizing the multinomial likelihood function in (\ref{hp2}) is less efficient than the ones based on individual observation, it is asymptotically efficient. Note that the group probability $P_{i} (\theta _{G} ,\theta _{F} )$ in (\ref{hp1}) can be obtained by first evaluating the cdf of a beta random variable at $F(x_{i-1} ;\theta _{F} )$ and $F(x_{i} ;\theta _{F} )$and then computing the difference between the two values. This reduces the complexity of programming required to calculate the integrations, because algorithms for evaluations of cdf are available readily in most statistical software.

 Next, the first derivative $L(\theta _{G} ,\theta _{F} )$ will be presented. Let $\Theta =$($\theta _{G} $,$\theta _{F} $)\textit{T}. The first derivative of $L(\Theta )$with respect to $\Theta $are
\begin{equation}
\frac{dL(\Theta )}{d\Theta } =\sum _{i=1}^{r}\frac{n_{i} }{P_{i} (\Theta )}  \frac{dP_{i} (\Theta )}{d\Theta }  \label{hp3}
\end{equation}
Note that the parameter vector$\theta _{F} $ are the parameters involved in the function \textit{F}. The derivatives of $P_{i} (\theta _{G} ,\theta _{F} )$in (\ref{hp1}) with respect to $\theta _{G} $ and $\theta _{F} $ are given by

\[\frac{dP_{i} (\theta _{G} ,\theta _{F} )}{d\theta _{G} } =\int _{F(x_{i-1} ;\theta _{F} )}^{F(x_{i} ;\theta _{F} )}\frac{dg(t;\theta _{G} ,\theta _{F} )}{d\theta _{G} }  dt;\]
\[\frac{dg(t;\theta _{g} ,\theta _{F} )}{d\alpha } =g(t)\left[-\frac{d\log B(\alpha ,\beta )}{d\alpha } +\ln t\right];\]
\[\frac{dg(t;\theta _{g} )}{d\beta } =g(t)\left[-\frac{d\log B(\alpha ,\beta )}{d\beta } +\ln (1-t)\right]; \]
\begin{equation}
 \frac{dP_{i} (\theta _{G} ,\theta _{F} )}{d\theta _{F} } =g\left[F(x_{i} ;\theta _{F} );\theta _{G} \right]\frac{dF(x_{i} ;\theta _{F} )}{d\theta _{F} } -g\left[F(x_{i-1} ;\theta _{F} );\theta _{G} ,\theta _{F} \right]\frac{dF(x_{i-1} ;\theta _{F} )}{d\theta _{F} }; \label{hp4}
\end{equation}
and $d\log B(\alpha ,\beta )/d\alpha $=$\Psi (\alpha )-\Psi (\alpha +\beta ).$  The nonlinear optimization subroutines in SAS can be employed by specifying the equation in (\ref{hp2})to be maximized and the gradient function in (\ref{hp3}).  Both the likelihood function to be maximized and the gradient function vary with the distribution functions$F(\cdot )$ under consideration, and the resulting functional forms of (\ref{hp4}
)  can be tedious and therefore not presented for any $F(\cdot )$under consideration here.

\section{Estimation and Comparisons}

The nonlinear Newton-Raphson method in SAS was employed with the
specification of the function to be minimized and the corresponding
gradient function. The income data were in a group format and can be
found on the Census Bureau's web site.  The first group consists of
families making less than $\$$25,000, and the last group of  more than
$\$$250,000.  In the evaluation of (\ref{hp1}) and (\ref{hp2}), the value
of the cdf $F(\cdot )$is set to be 0 at the lower boundary of the first
class and 1 at the upper boundary of the last class in our SAS
programs.

The results for years 2003, 2004 and 2005 are reported in Tables 2, 3
and 4.  The mean ($\mu_i$) and frequency ($n_i$) for each group are
reported on the Census Bureau's web site and the approximated mean
income for each year can be calculated by $\sum n_i \mu_i/\sum n_i$.
The approximated sample mean incomes (in $\$$10,000) for 2003, 2004 and
2005 are 6.598, 6.140 and 7.040, respectively.  Note that year 2004 has
the lowest means among the three years. The estimated means in the
following tables are calculated using the estimated parameter values in
the mean expressions given Table 1. The resulting estimated means using
the skewed-t appear to overestimate.  The sum of squared errors (SSE)
between the relative frequency $n_{i} /N$and the estimated frequency
$\hat{P}_{i} (\Theta )$ or the absolute errors (SAE), and chi-square
$\chi ^{2}$ are also reported.

The generalized four-parameter log-F distribution appears to yield the best fit in terms of chi-squares and SAE, and the generalized beta of the second type (GB2) in terms of SSE. Overall, the log-F performs well which is consistent with Jones' belief that Log-F provides the most tractable instances of families with power and exponential tails.  The two-parameter skew-\textit{t} performs relatively poor in the results. As in McDonald (1984), the generalized beta of the second type provides better fit than the generalized beta of the first type (GB1).  Trailing behind the log-F and GB2 is the beta-Weibull.  The three-parameter beta-exponential and beta-Weibull provide better fit than the GB1 in terms of all measures of goodness fit. Thought the skew-t has second worst performance, it appears to perform much better than beta-normal.  The beta-normal distribution noticeably performs the worst. Note that the normal distribution itself is a poor fit for skewed data.

\begin{table*}
\caption{Estimated Results for 2003}
\begin{tabular}{|l|r|r|r|r|r|r|r|}\hline
 & {|}{GB1} & {GB2} & {BN} & {skew-t} &  {Log-F} & {BE} & {BW} \\ \hline
$\alpha$   & 1.955     & 0.490 & 2.348 & 7.822 & 3.468 & 1.700 & 1.748 \\ \hline
$\beta $   & 2830.689  & 1.111 & 0.369 & 0.936 & 0.195 & 0.799 & 0.875 \\ \hline
\textit{a} & 0.889     & 2.724 & 0.000 &       & 2.256 & 0.257 & 0.251 \\ \hline
\textit{b} & 22685.000 & 8.297 & 4.012 &       & 2.294 &       & 0.982 \\ \hline
Est. Means      & 6.517& 6.618& 6.642& 7.474& 6.017& 6.498 & 6.485\\ \hline
1000 *SSE       & 0.929    & 0.504& 9.390& 2.225 & 0.562 & 0.886 & 0.877  \\ \hline
SAE        & 0.161     & 0.123 & 0.379 & 0.224 & 0.125 & 0.158 & 0.158 \\ \hline
$\chi ^{2}$& 3816.534  &1972.524 & 65394.750 & 6103.421 & 1926.019 & 3698.846 & 3707.640 \\ \hline
\end{tabular}
\end{table*}

\begin{table*}
\caption{Estimated Results for 2004}
\begin{tabular}{|l|r|r|r|r|r|r|r|}\hline
 &{GB1} & {GB2} & {BN} & {skew-t} &  {Log-F} & {BE} & {BW} \\ \hline
$\alpha $ & 1.680 & 0.485 & 2.328 & 7.649 & 3.226 & 1.608 & 1.552 \\ \hline
$\beta $ & 6558.823 & 1.179 & 0.342 & 0.894 & 0.187 & 0.948 & 0.829 \\ \hline
\textit{a} & 0.951 & 2.647 & 0.000 &  & 2.043 & 0.209 & 0.219 \\ \hline
\textit{b} & 39366.000 & 8.951 & 3.942 &  & 2.039 &  & 1.024 \\ \hline
Est. Means & 6.679 & 6.784 & 6.818 &      7.883 & 6.075& 6.681 & 6.684 \\ \hline
1000* SSE & 0.841   & 0.477 & 8.697 & 2.727 & 0.420 & 0.789 & 0.799 \\ \hline
SAE & 0.151 & 0.122 & 0.366 & 0.232 & 0.106 & 0.148 & 0.148 \\ \hline
$\chi ^{2} $ & 3501.860 & 1922.756 & 55209.277 & 6851.514 & 1451.627 & 3449.960 & 3428.916 \\ \hline
\end{tabular}
\end{table*}

\begin{table*}
\caption{Estimated Results for 2005}
\begin{tabular}{|l|r|r|r|r|r|r|r|}\hline
 &{GB1} & {GB2} & {BN} & {skew-t} &  {Log-F} & {BE} & {BW} \\ \hline
$\alpha $ & 1.672 & 0.449 & 2.311 & 7.734         & 3.254 & 1.609 & 1.547 \\ \hline
$\beta $ & 3945.308 & 1.041 & 0.330 & 0.885         & 0.170 & 0.950 & 0.822 \\ \hline
\textit{a} & 0.954 & 2.838 & 0.000 &  & 0.003 & 0.205 & 0.215 \\ \hline
\textit{b} & 23068.000 & 8.773 & 3.937 &  & 0.942 &  & 1.027 \\ \hline
Est.  Means & 6.811 & 7.145 & 6.931 &       8.116 & 6.256 &       6.801 & 6.801 \\ \hline
1000*SSE & 0.859& 0.429 & 9.002e  & 2.789  & 0.446  & 0.799\newline  & 0.812 \\ \hline
SAE & 0.151 & 0.114 & 0.364 & 0.237   & 0.105 & 0.147 & 0.148 \\ \hline
$\chi ^{2} $ & 3529.821 & 1764.919 & 52066.738 & 7096.854 & 1525.312 & 3475.881 & 3452.047 \\ \hline
\end{tabular}
\end{table*}

Next, in order to have a better picture on how the tail for each distribution fitted to the data, the estimated density functions based on the 2005 income data are presented in the following graph. The skewed-t appears to result a thicker tail than others.

In summary, the log-F provides the best relative fit and then followed by the generalized beta of the second type. Among other distributions in the family of the generalized beta distribution that were fit to the data, the beta-normal appears to perform poorly. The two-parameter skew-t distribution can probably extended to four-parameter one whose mathematical properties including moments and shapes needs further studied.

\begin{landscape}
\begin{figure}
\psfig{file=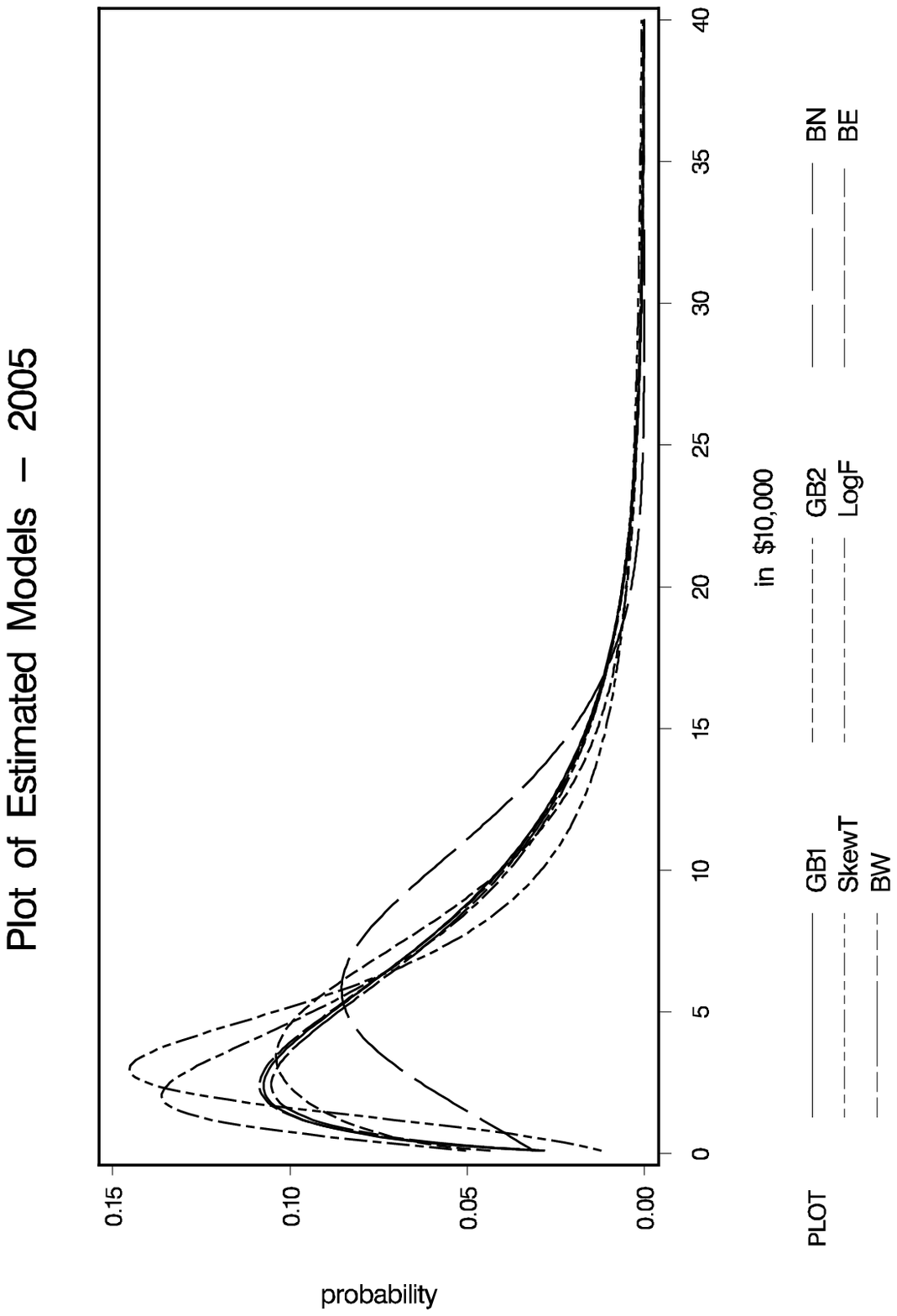}
%
\end{figure}
\end{landscape}


\begin{thebibliography}{20}

\bibitem{}
Brown, B.W, M.S. Floyd, and L.B. Levy, (2002). The log-F: a distribution for all seasons. \textit{Computational Statistics}, 17, 47-58.


\bibitem{r2}
Cardeno, L., D. K. Nagar, and L. E. Sanchez, (2005).  Beta Type 3 Distribution and Its Multivariate Generalization. \textit{Journal of Mathematical Sciences} 21, 225-241.


\bibitem{r3}
Eugene, N., C. Lee, and F. Famoye, (2002). Beta-normal Distribution and its applications. \textit{Communication in Statistics -- Theory and Methods}, 31: 497-512.


\bibitem{r4}
Famoye, F., Lee, C. \& Eugene, N. (2004). Beta-Normal Distribution: Bimodality Properties and Applications. \textit{Journal of Modern Applied Statistical Methods}, Vol. 3, 85-103.


\bibitem{r5}
Famoye, F. , Lee, C. \& Olugbenga Olumolade, (2005). The Beta-weibull distribution. \textit{Journal of Statistical Theory and Applications,} 121-138. \textit{  }


\bibitem{r6}
Ferreira, Jose T.A.S. and Mark F.J. Steel, (2004). A Constructive Representation of Univariate Skewed Distributions.. Econometrics 0403002, Economics Working Paper Archive EconWPA.


\bibitem{r7}
Gradshteyn, I.S. and placeI. M. Ryzhik. (2000). \textit{Table of Integrals, Series and Products}. Academic Press: San Diego.


\bibitem{r8}
Gupta, A. K.  and S.Nadarajah, (2004). On the moments of the beta normal distribution. \textit{Communication in Statistics -- Theory and Methods}, 31: 1-13.


\bibitem{r9}
Jones, M. C. (2001). A skew-t distribution, In country-regionC.A. Charalambides, M. V. Kourtras, and placeN. Balarishnan, eds,. \textit{Probability and Statistical Models with Applications}, pp. 269-278. Chapman and Hall, London.


\bibitem{r10}
Jones, M. C, and M. J. Faddy, (2003).  A skew extension of the \textit{t }distribution, with applications. \textit{Journal of the Royal Statistical Society}, Series B, 65:159-174.


\bibitem{r11}
Jones, M. C. (2004). Family of Distributions Arising from Distribution of Order Statistics. \textit{Test} Vol. 13, No. 1, pp.1-43.


\bibitem{r12}
Kalbfleish, J. D. and R.L Prentice, (1980). The \textit{Statistical Analysis of Failure Time Data}. John Wiley and Sons, StateplaceNew York.


\bibitem{r13}
McDonald, J. B. (1984). Some generalization functions for the size distribution of income. \textit{Econometrica}, 52, 3, 647-663.


\bibitem{r14}
Nadarajah, S. (2005). Exponentiated beta distributions. {\it Computers \& Mathematics with Application}.  49, 1029-1035.


\bibitem{r15}
Nadarajah S. and placeS. Kotz, (2004). The beta Gumbel Distribution. \textit{Mathematical Problems in Engineering, }1004:4, 323-332.


\bibitem{r16}
Singh, K.P., C.M. Lee, and E.O. George, (1988). On generalized log-logistic model for censored survival data. \textit{Biometrical Journal}, 30, 843-850.


\bibitem{r17}
Thurow, L. C. (1970). Analyzing the American Income Distribution. \textit{Papers and proceedings}, \textit{American Economics Association}, 60, 261-269.


\end{thebibliography}
\end{document}